\documentclass[a4paper,dvipdfmx]{article}
\usepackage{amssymb,amsmath}
\usepackage{amsthm}
\usepackage{bm}%
\usepackage{stmaryrd}
\usepackage{cite}

\def\nn{\nonumber}

\def\hf{\frac{1}{2}}

\def\Zbb{{\mathbb Z}}
\def\Zgr{$ {\mathbb Z}_2$}
\def\Z2{${\mathbb Z}_2 \times {\mathbb Z}_2$}
\def\g{\mathfrak{g}}
\def\gc{\hat{{\mathfrak g}}_{\ell_1,\ell_2}}

\def\CA{\Z2-graded color superalgebra }

\newcounter{thMM}
\newcounter{leMM}
\newcounter{deFF}
\newcounter{exMP}
\newcounter{prMM}
\newcounter{coro}
\newcounter{rema}
\newenvironment{theorem}[1]{\refstepcounter{thMM}\trivlist
   \item[\hskip19pt{\sc #1~\arabic{thMM}.}]\it\hskip3pt}{\endtrivlist}
\newenvironment{lemma}[1]{\refstepcounter{leMM}\trivlist
   \item[\hskip19pt{\sc #1~\arabic{leMM}.}]\it\hskip3pt}{\endtrivlist}
\newenvironment{definition}[1]{\refstepcounter{deFF}\trivlist
   \item[\hskip19pt{\sc #1~\arabic{deFF}.}]\rm\hskip3pt}{\endtrivlist}

\begin{document}
%
%
%
%
\thispagestyle{empty}

\title{ \large $\mathbb{Z}_2 \times \mathbb{Z}_2$ generalizations of infinite dimensional Lie superalgebra of conformal type with complete classification of central extensions }
\author{ N. Aizawa \\   
Department of Physical Science, Graduate School of Science, \\
Osaka Prefecture University, Nakamozu Campus, \\
Sakai, Osaka 599-8531 Japan \\
   aizawa@p.s.osakafu-u.ac.jp 
      \\[2ex]
    P. S. Isaac \\
    School of Mathematics and Physics, \\
    The University of Queensland, St Lucia QLD 4072, Australia
    \\
    psi@maths.uq.edu.au
      \\[2ex]
    J. Segar \\
    Department of Physics, \\
    Ramakrishna Mission Vivekananda College,  \\ 
    Mylapore, Chennai 600 004, India
     \\ 
     segar@rkmvc.ac.in  }

\maketitle

\vfill
\begin{abstract}

We introduce a class of novel $\mathbb{Z}_2 \times \mathbb{Z}_2$-graded color superalgebras of infinite dimension. 
It is done by realizing each member of the class in the universal enveloping algebra of a Lie superalgebra which is a module extension of the Virasoro algebra. 
Then the complete classification of central extensions of the $\mathbb{Z}_2 \times
\mathbb{Z}_2$-graded color superalgebras is presented. It turns out that infinitely many members of the class have non-trivial extensions. We also demonstrate that the color superalgebras (with and without central extensions) have adjoint and superadjoint operations.
\end{abstract}


\clearpage
\setcounter{page}{1}
%
%
%
\section{Introduction}

Continuous symmetry is one of the most fundamental notions in mathematics and theoretical physics. 
In many cases a Lie group is used to describe continuous symmetries and its Lie algebra corresponds to the infinitesimal transformations.  
Two types of generalizations of Lie groups and Lie algebras are also used to describe enlarged notions of 
symmetry. 
The first type involves the supergroups which are defined by introducing a $
\Zbb_2 $ grading structure on Lie groups. 
In physics, this provides transformations combining bosonic and fermionic degrees of freedom. 
The second type relates to the quantum groups which are a type of deformation of the coordinate ring of a Lie group.
A quantum group may describe ``continuous" transformations in a non-commutative space. 
It is widely recognized that these two generalizations play rather important roles, not only in the
context of symmetry, but also in various fields of mathematics and physics.

These two generalizations, however, are not the only ones extending Lie groups and Lie algebras. 
Indeed, a generalization of Lie algebras was considered by Ree in 1960 \cite{Ree} and then
rediscovered, together with a corresponding generalization of Lie groups in the late 1970s \cite{rw1,rw2,sch}. 
The basic idea of the generalization is to introduce a new grading structure to the underlying
vector space of the Lie algebra.  
In other words, it replaces the $\mathbb{Z}_2 $ grading of superalgebras with a more general abelian group. 
Such a generalization is called \textit{color} (super)algebras after the work trying to unify the
spacetime and internal symmetries \cite{lr}.
Like supergroups and quantum groups, the color (super)groups will also provide a new enlarged symmetry. 

Of particular note are the studies that have been motivated by applications of the color
(super)algebras with abelian group \Z2. Such works consider, for example, the enlarged symmetries in physical problems in the context of
extensions of spacetime symmetries (such as the de-Sitter and Poincar\'e algebras) in supergravity theory,
quasispin and parastatistics \cite{vas,jyw,zhe,Toro1,Toro2,tol2,tol}.
Furthermore, these color (super)algebras have recently been
recognized for their potential physical application in supersymmetric and
superconformal quantum mechanics \cite{Bru,BruDup,NaAmaDoi,Ueba} or double-field theory \cite{BruIbar}. 
It is our view, therefore, that these
particular color (super)algebras (i.e. with \Z2 grading) warrant further scrutiny.

Recently, it was observed that symmetries of a partial differential equation, called the
L\'evy-Leblond equation, are given by a color superalgebra having a \Z2 grading structure (we call
it \CA throughout this paper) \cite{aktt1,aktt2}.
The L\'evy-Leblond equation is a quantum mechanical wave equation of a spin 1/2 particle in a
non-relativistic setting so that it may be viewed as one of the fundamental equations in physics.
In fact, the color superalgebra defining the symmetry of the L\'evy-Leblond equation is realized in
the universal enveloping algebra (UEA) of the super Schr\"odinger algebra.
Furthermore, it is shown that some \CA can be realized in the UEA of the superconformal Galilei
algebras which contain the super Schr\"odinger algebras as a special case \cite{NAJS,NaPsiJs}. 
It is also known that the Clifford algebras are a color superalgebra of $\mathbb{Z}_2^{\otimes n}$
grading \cite{NA2}.

Since the work of Bargmann \cite{Barg}, the utility of central extensions has been understood for
their effectiveness in dealing with projective representations at the level of the group, which  
occur somewhat naturally in the context of quantum field theory. The Virasoro algebra is arguably
the most well-studied example of a Lie algebra (an infinite-dimensional one) that occurs as a central
extension of another Lie algebra (the Witt algebra), describing conformal algebraic symmetries in
the plane. The motivation to study color generalizations of this algebra, as we have done in this
article, is therefore worthwhile, and has been inspired by the mathematical works
\cite{OvRo1,OvRo2,Mar,MaOvRo} and papers related to gravitation \cite{BMS4,Sachs}.

One could easily imagine that color (super)algebras are not at all unusual in the context of
(super)symmetries of physical systems. 
In order to provide an algebraic foundation for such symmetry 
and an extension of \cite{NAJS,NaPsiJs} (where finite dimensional algebras were considered),  
we provide a class of infinite dimensional \CA in the
present work. Each member of the class is realized in the UEA of a Lie superalgebra which is a
module extension of the Virasoro algebra. As a result these color superalgebras also have the
Virasoro algebra as a subalgebra.
We then give the complete classification of central extensions of this class of color superalgebras. 
Namely, we identify many members of the class that give rise to various non-trivial extensions. 
In this way, we introduce a class of novel infinite dimensional \CA which has a close relation to Lie superalgebras. 

Various mathematical studies of the algebraic and geometric aspects of the color (super)algebras have been
undertaken since their introduction, see e.g., 
\cite{GrJa,sch2,sch3,SchZha,Sil,ChSiVO,SigSil,PionSil,CART,
NAJS,NA1,NA2,StoVDJ,NaPsiJs,CGP1,CGP2,CGP3,CGP4,Pon,CoKPo,BruPon,BruPon2,MohSal,Bru}
and references therein. 
Here we mention the motivation considering the module extensions of the Virasoro algebra. 
As we have already mentioned above, such algebras have been studied from a mathematical perspective 
\cite{OvRo1,OvRo2,Mar,MaOvRo} and in the context of gravitational physics \cite{BMS4,Sachs}.
More recently, these algebras have attracted some interest in
physics in relation to 
AdS/CFT correspondence
for example (see \cite{NAJS2} for more detailed references).
Therefore, \Z2 graded versions of such algebras may have potential applications to physical problems. 

We plan the paper as follows. 
In the next section, we give the definition of a \CA which is the only color generalization considered in the present work. 
We also give a class of Lie superalgebras, denoted by $\mathfrak{v}_{\ell_1,\ell_2}, $  which is a module extension of the Virasoro algebra and our starting point of construction of color superalgebras. 
We collect the results of the present work in \S \ref{SEC:Main} 
The first result is a construction of a \CA by  realizing it in the UEA of $\mathfrak{v}_{\ell_1,\ell_2}. $ 
The \CA so constructed has central extensions. 
The second result is the complete classification of  all possible central extensions. 
The method employed is elementary but the lengthy discussion is needed. 
So the detail is presented in \S \ref{SEC:Proof}  
The third result relates to adjoint and superadjoint operations in the \Z2-graded color superalgebra. 
We introduce adjoint and superadjoint operations to the \Z2-graded setting which are a natural generalization of the $\mathbb{Z}_2$-graded case. 
We then demonstrate that our \CA do indeed have these operations by giving them explicitly. 

Throughout this article, all vector spaces are considered over the ground field $\mathbb{C}. $

%
%
%
\section{Preliminaries}

\subsection{\Z2-graded color superalgebra}

Here we give the definition of a \Z2-graded color superalgebra \cite{rw1,rw2}. 
Let $ \g $ be a vector space  and $ \bm{a} = (a_1, a_2)$ an element of  \Z2. 
Suppose that $ \g $ is a direct sum of graded components:
\begin{equation}
   \g = \bigoplus_{\bm{a}} \g_{\bm{a}} = \g_{(0,0)} \oplus \g_{(0,1)} \oplus \g_{(1,0)} \oplus \g_{(1,1)}.
\end{equation}
In what follows, we denote homogeneous elements of $ \g_{\bm{a}} $ as $ X_{\bm{a}}, Y_{\bm{a}},
Z_{\bm{a}}$. 
If $\g$ admits a bilinear operation (the general Lie bracket), denoted by $ \llbracket \cdot, \cdot \rrbracket, $ 
satisfying the identities
\begin{align}
  & \llbracket X_{\bm{a}}, Y_{\bm{b}} \rrbracket \in \g_{\bm{a}+\bm{b}}
  \\[3pt]
  & \llbracket X_{\bm{a}}, Y_{\bm{b}} \rrbracket = -(-1)^{\bm{a}\cdot \bm{b}} \llbracket Y_{\bm{b}}, X_{\bm{a}} \rrbracket,
  \\[3pt]
  & (-1)^{\bm{a}\cdot\bm{c}} \llbracket X_{\bm{a}}, \llbracket Y_{\bm{b}}, Z_{\bm{c}} \rrbracket \rrbracket
    + (-1)^{\bm{b}\cdot\bm{a}} \llbracket Y_{\bm{b}}, \llbracket Z_{\bm{c}}, X_{\bm{a}} \rrbracket \rrbracket
    + (-1)^{\bm{c}\cdot\bm{b}} \llbracket Z_{\bm{c}}, \llbracket X_{\bm{a}}, Y_{\bm{b}} \rrbracket
\rrbracket =0, 
    \label{gradedJ}
\end{align}
where
\begin{equation}
  \bm{a} + \bm{b} = (a_1+b_1, a_2+b_2) \in {\mathbb Z}_2 \times {\mathbb Z}_2, \qquad \bm{a}\cdot \bm{b} = a_1 b_1 + a_2 b_2\in {\mathbb Z}_2,
\end{equation}
then $\g$ is referred to as a \Z2-graded color superalgebra. 

The enveloping algebra of $\g$ is the \Z2-graded unital associative algebra with relations
\begin{equation}
  X_{\bm{a}} Y_{\bm{b}} - (-1)^{\bm{a}\cdot \bm{b}} Y_{\bm{b}} X_{\bm{a}} = \llbracket X_{\bm{a}}, Y_{\bm{b}} \rrbracket.
  \label{gradedcom}
\end{equation}
One key observation is that for homogeneous elements, their general Lie bracket will coincide with either a commutator $(\bm{a}\cdot \bm{b}=0)$ or anticommutator $(\bm{a}\cdot \bm{b}=1)$ .  
We also use the notation $ [X_{\bm{a}}, Y_{\bm{b}}] $ (in case $ \bm{a}\cdot \bm{b}=0$) and $ \{ X_{\bm{a}}, Y_{\bm{b}}  \} $ (in case $ \bm{a}\cdot \bm{b}=1$) for the general Lie bracket in order to emphasize that given elements commute or anticommute.
%
%
It should be noted that  $ \g_{(0,0)} \oplus \g_{(0,1)} $ and $ \g_{(0,0)} \oplus \g_{(1,0)} $ are
subalgebras of $\g$ (with \Z2-grading).
We remark that this is a natural generalization of Lie superalgebra which is defined with a \Zgr-graded structure:
\begin{equation}
  \g =  \bigoplus_{\bm{a}} \g_{\bm{a}} = \g_{(0)} \oplus \g_{(1)},
\end{equation}
instead with 
\begin{equation}
  \bm{a} + \bm{b} = (a+b)\in {\mathbb Z}_2, \qquad \bm{a} \cdot \bm{b} = ab\in {\mathbb Z}_2.
\end{equation}

The central element of a color superalgebra is defined as an element having vanishing general Lie  bracket with any elements of $\mathfrak{g}_{\ell_1,\ell_2}.$ 
Namely, if $C$ is an  centeral element, then $ \llbracket C, X \rrbracket = 0 $ for $\forall X \in \mathfrak{g}_{\ell_1,\ell_2}.$ In other words, the central element $C$ also has the \Z2 degree.

\subsection{A module extension of Virasoro algebra }

In this section a class of infinite dimensional Lie superalgebras of the standard $ \mathbb{Z}_2$-grading, which is the basis of our construction of color superalgebra, is introduced. 
This is done by extending the Virasoro algebra with its representations \cite{Schtt,ChaPre,GKO,FNZ}.

We start with the Witt algebra W which is the Lie algebra of the group of the orientation preserving diffeomorphisms of $ S^1.$   
The algebra coincides with the algebra of vector fields on $S^1.$  
Let $ L_f := f(z) \frac{d}{dz}$ be a vector field where $ f(z+2\pi) = f(z),$ then we have the commutator for $W:$
\begin{equation}
    [ L_f, L_g] = \big(f(z) g'(z) - f'(z) g(z) \big) \frac{d}{dz}.  
\end{equation} 
This may equivalently be written in terms of the  `Fourier components' $ L_m = -z^{m+1} \frac{d}{dz}$ as $ [L_m, L_n] = (m-n) L_{m+n}.$ 
The Virasoro  algebra Vir is the unique (up to isomorphism) central extension of the Witt algebra.  The central extension is given by the Gelfand-Fuchs cocycle \cite{GFcocy}
\begin{equation}
  c(L_f, L_g) = \int_{S^1} f'(z) g''(z) dz,
\end{equation}  
or in Fourier components
\begin{equation}
   c(L_m, L_n) = m(m^2-1) \delta_{m+n,0}.
\end{equation}
The defining relations of Vir are given explicitly in terms of the Fourier components by
\begin{equation}
  [L_m, L_n] = (m-n) L_{m+n} + \frac{c}{12}m(m^2-1) \delta_{m+n,0}. 
  \label{VirDef}
\end{equation}

  Now we consider unitary representations of Vir. 
It is known that there exist two essentially different families of irreducible unitary representations of Vir, i.e., the highest weight representations and the space of  tensor densities on $ S^1$ \cite{ChaPre,GKO}. 
Here we consider the latter. 
Let $ {\cal F}_{\lambda}$ be the space of all tensor densities on $ S^1$ of degree $\lambda:  a = a(z) dz^{\lambda}.$ 
The Witt algebra acts on $ {\cal F}_{\lambda}$ by the Lie derivative:
\begin{equation}
   L_f.a = \big(f(z) a'(z) + \lambda f'(z) a(z) \big) dz^{\lambda}
\end{equation}
and the action extends to Vir by setting $ c.a = 0.$ 
Recalling that Vir has the subalgebra $ sl(2) = \langle\; L_{\pm 1}, L_0 \; \rangle,$ we take $ \lambda = -\ell $ where $\ell$ takes a spin value: $ \ell = 0, \frac{1}{2}, 1, \frac{3}{2}, 2, \dots. $ 
In the basis $ a_n = z^n dz^{-\ell}, \ n \in \mathbb{Z},$  one gets $ L_m.a_n = ( (m+1) \ell - n) a_{m+n}. $ 
We express this relation, by introducing the spin variable $r := n-\ell \in \mathbb{Z}+\ell $ as $ L_m.a_r = (m \ell - r) a_{m+r}. $
%
It follows that the semi-direct sum  Vir $\ltimes {\cal F}_{-\ell}$ may be understood as an extension of Vir  with the additional relations $ [L_m, a_r ] = (m\ell-r) a_{m+r}$ and $ [c, a_r] = 0 $  
(c.f., \cite{OvRo1,OvRo2,Mar} where  extensions of Vir by $ {\cal F}_{\lambda}$ which are not isomorphic to the semi-direct sum are classified).  

 We are now ready to define our ($\mathbb{Z}_2$-graded) Lie superalgebra. 
Consider two Vir modules $ {\cal F}_{-\ell_1} $ and $ {\cal F}_{-\ell_2} $ where $\ell_1 $ and $ \ell_2$ take a spin value. 
We denote a basis of $ {\cal F}_{-\ell_1} $ (resp. $ {\cal F}_{-\ell_2} $) by $ P_r $ (resp. $X_u$) with $ r \in \mathbb{Z}+ \ell_1$ (resp. $ u \in \mathbb{Z}+\ell_2$). 
Then the Lie superalgebra, denoted by $\mathfrak{v}_{\ell_1,\ell_2},$ is defined as a super abelian extension of Vir by $ {\cal F}_{-\ell_1} \oplus \Pi{\cal F}_{-\ell_2}$ where $ \Pi{\cal F}_{-\ell_2}$ indicates that ${\cal F}_{-\ell_2}$ is considered as an odd vector space in the sense of $\mathbb{Z}_2$-grading. 
The defining relations of $\mathfrak{v}_{\ell_1,\ell_2}$ are given explicitly by \eqref{VirDef} and 
\begin{align}
    [L_m, P_r] &= (m \ell_1-r) P_{m+r}, \nn \\
    [L_m, X_u] &= (m \ell_2-u) X_{m+u}, \nn \\
    [P_r, P_s] &= [P_r, X_u] = \{ X_u, X_v \} = 0, \nn \\
    [L_m, c] &= [P_r, c] = [X_u, c] = 0.  \label{super-def}
\end{align}

Throughout this article the indices $ m, n $ are reserved to indicate integers ($m, n \in \Zbb$), while 
 $ r, s, t$ (resp. $u, v, w$) are reserved to indicate numbers taking values in $ \Zbb +
\ell_1$ (resp. $\Zbb+\ell_2$).

%
%
%
%
\section{Algebraic structure} \label{SEC:Main}

In this section, we present the main results of the current work. 
We denote the universal enveloping algebra of $\mathfrak{v}_{\ell_1,\ell_2} $  by $U(\mathfrak{v}_{\ell_1,\ell_2}).$

\subsection{\Z2-graded color superalgebra realized in $U(\mathfrak{v}_{\ell_1,\ell_2})$}

Our first result is the existence of the \Z2-graded color superalgebra which is realized in the $\mathbb{Z}_2$-graded algebra $U(\mathfrak{v}_{\ell_1,\ell_2}).$  
To see this, we take particular elements in $U(\mathfrak{v}_{\ell_1,\ell_2}):$
\begin{equation}
 P_{rs} = 2P_r P_s, \qquad 
 X_{uv} = 2X_u X_v, \qquad
 T_{ru} =2P_r X_u  \label{newele}
\end{equation}
and consider the infinite dimensional $ \mathbb{Z}_2$-graded vector space $ \g_{\ell_1,\ell_2} $   spanned by the elements
\begin{equation}
  (0) \quad L_m, \quad P_r, \quad P_{rs}, \quad X_{uv}, \quad c,  \qquad (1) \quad X_u, \quad T_{ru}
\end{equation}
where the $\mathbb{Z}_2$-degree is indicated in parentheses and the indices run over all allowed values, as described previously. 
The element $ P_{rs} $ is symmetric in its indices, $ X_{uv} $ is antisymmetric, and $ T_{ru} $ has no particular symmetry. 
Obviously,
\begin{equation}
   P_{rs} = \{ P_r, P_s \}, \qquad 
   X_{uv} = [ X_u, X_v], \qquad
   T_{ru} = \{ P_r, X_u \}.  
\end{equation}
This is a wrong choice of commutator and anticommutator in the $\mathbb{Z}_2$-graded setting. 
However, it is correct if we consider that $ P_r $ and $ X_u$ have the \Z2-degree $(0,1) $ and $(1,1),$ respectively. 
This observation leads us to assign the \Z2-degree to the elements of $ \g_{\ell_1,\ell_2} $ as follows
\begin{equation}
  \begin{array}{cl}
     (0,0) & \ L_m, \quad P_{rs}, \quad X_{uv}, \quad c \\
     (0,1) & \ P_r \\
     (1,0) & \ T_{ru} \\
     (1,1) & \ X_u
  \end{array}
  \label{color-deg}
\end{equation}
By this assignment, we may view the enveloping algebra $U(\mathfrak{v}_{\ell_1,\ell_2})$ as a \Z2-graded algebra as well as $\mathbb{Z}_2$-graded one. 
The enveloping algebra $U(\mathfrak{v}_{\ell_1,\ell_2})$ is a quotient of the tensor algebra  $T(\mathfrak{v}_{\ell_1,\ell_2})$ on the vector space $\mathfrak{v}_{\ell_1,\ell_2}$ by its ideal $I$ generated by the relations \eqref{VirDef} and \eqref{super-def}. 
Since \eqref{VirDef} and \eqref{super-def} are homogeneous with respect to the \Z2 degree given in \eqref{color-deg}, both the ideal $I$ and $T(\mathfrak{v}_{\ell_1,\ell_2})$ are viewed as \Z2-graded algebra so that the quotient algebra is also \Z2-graded. 
It follows that if the vector space $\g_{\ell_1,\ell_2}$ is closed under the \Z2 version of commutator and anticommutator, then it will be a color superalgebra at the abstract level. One may easily verify that this is indeed the case.  For illustrative purpose we give some examples of the computation of the relations.
\begin{equation}
  \{ X_u, T_{rv} \} \stackrel{\eqref{newele}}{=} \{X_u, 2P_r X_v \} 
   = 2 [X_u, P_r] X_v + 2 P_r \{ X_u, X_v \} 
   \stackrel{\eqref{super-def}}{=} 0.  
\end{equation}
Similarly, one may verify that
\begin{equation}
  [P_r, P_{st} ] = [P_r, X_{uv}] = [T_r, T_{su}] = [X_u, P_{rs}] = [X_u, X_{vw}] = 0
\end{equation}
which shows that higher order products of $P_r, X_u$ are not necessary. 
\begin{align}
  [L_m, P_{rs}] &\stackrel{\eqref{newele}}{=} 2[ L_m, P_r P_s] 
    = 2[L_m, P_r] P_s + 2 P_r [L_m, P_s]
    \nonumber \\
    &\stackrel{\eqref{super-def}}{=} (m \ell_1 -r) 2P_{m+r} P_s + (m \ell_1-s) 2P_r P_{m+s}
    \nonumber \\
    & \stackrel{\eqref{newele}}{=} (m\ell_1-r) P_{m+r,s} + (m\ell_1-s) P_{r,m+s}.
\end{align}
Repeating the computation in this way, one may find that the non-vanishing \Z2 brackets for $ \g_{\ell_1,\ell_2} $  are given by
\begin{align}
   [L_m, L_n] &= (m-n) L_{m+n} + \frac{c}{12}m(m^2-1) \delta_{m+n,0}, \nn \\
   [L_m, P_r] &= (m\ell_1-r) P_{m+r}, \nn \\
   [L_m, X_u] &= (m\ell_2-u) X_{m+u}, \nn \\
  [L_m, P_{rs}] &= (m\ell_1-r) P_{m+r,s} + (m\ell_1-s) P_{r,m+s},  \nn \\
  [L_m, X_{uv}] &= (m\ell_2-u) X_{m+u,v} + (m\ell_2-v) X_{u,m+v},  \nn \\
  [L_m, T_{ru}] &= (m\ell_1-r) T_{m+r,u} + (m\ell_2-u) T_{r,m+u},  \nn \\
  \{ P_r, P_s \} &= P_{rs}, \quad [X_u, X_v] = X_{uv}, \quad \{P_r, X_u\} = T_{ru}. 
  \label{Color-def}
\end{align}

Therefore, we have established

\begin{lemma}{Proposition}
The vector space $\mathfrak{g}_{\ell_1,\ell_2}$ with the non-vanishing relations \eqref{Color-def} define a \Z2-graded color superalgebra.
\end{lemma}

Since the \Z2-graded color superalgebra $\mathfrak{g}_{\ell_1,\ell_2}$ is realized inside the $U(\mathfrak{v}_{\ell_1,\ell_2}),$ 
any representation of $\mathfrak{v}_{\ell_1,\ell_2}$ is converted to a representation of  $\mathfrak{g}_{\ell_1,\ell_2}.$

\subsection{Central extensions}

Our second result is the complete classification of all possible central extensions of $\mathfrak{g}_{\ell_1,\ell_2}.$  
We denote the central extension of $\mathfrak{g}_{\ell_1,\ell_2}$  by $\gc. $
The classification of the central extensions is summarized in the following theorem:

\begin{theorem}{Theorem} \label{Thm}
  All possible central extensions of $\mathfrak{g}_{\ell_1,\ell_2} $ are given as follows:
\begin{alignat*}{2}
   [L_m, P_r] &= (m\ell_1-r) P_{m+r} + \delta_{m+r,0}\, p(m),  \\
   [L_m, X_u] &= (m\ell_2-u) X_{m+u} + \delta_{m+u,0}\, x(m),   \\ 
  [P_{rs}, P_t] &= \delta_{r+s+t,0} \, h(r,s), 
  & [X_{uv}, X_w] &= \delta_{u+v+w,0}\, \eta(u,v),
  \\
  [P_{rs}, X_u] &= \delta_{r+s+u,0} \, \rho(r,s), & 
  [X_{uv}, P_r] &= \delta_{u+v+r,0}\, q(u,v), \\
  [T_{ru}, P_s] &= \delta_{r+u+s,0}\, \kappa(r,s), & 
  \{T_{ru}, X_{v} \} &= \delta_{r+u+v,0} \,\zeta(u,v),
\end{alignat*}
where
\begin{align*}
 p(m) &= (\delta_{\ell_1 0}\, m^2  + \delta_{\ell_1 1}\, m^3 )\, c_p,
 \\
 x(m) &= (\delta_{\ell_2 0}\, m^2  + \delta_{\ell_2 1}\, m^3 )\, c_x,
 \\
 h(r,s) &= \delta_{\ell_1 0} (r+s) c_h, 
 \\
 \eta(u,v) &= \delta_{\ell_2 0} (u-v) c_{\eta}, 
 \\
 \rho(r,s) &= -2 \big( \delta_{\ell_1 0} \delta_{\ell_2 0} (r-s) + \delta_{\ell_1 \hf} \delta_{\ell_2 0} (r^2-s^2) \big) c_{\kappa_A},
 \\
 q(u,v) &= -2 \big( \delta_{\ell_1 0} \delta_{\ell_2 0} (u-v) + \delta_{\ell_1 0} \delta_{\ell_2 \hf} (u^2-v^2) \big) c_{\zeta_A},
 \\
 \kappa(r,s) &= \delta_{\ell_1 0} \delta_{\ell_2 0} \big( (r+s) c_{\kappa_S} + (r-s) c_{\kappa_A} \big) + \delta_{\ell_1 \hf} \delta_{\ell_2 0} (r^2-s^2) c_{\kappa_A} + \delta_{\ell_1 0} \delta_{\ell_2 1}\, rs\, c_{\kappa_S},
 \\
 \zeta(u,v) &= \delta_{\ell_1 0} \delta_{\ell_2 0} \big( (u+v) c_{\zeta_S} + (u-v) c_{\zeta_A} \big) + \delta_{\ell_1 0} \delta_{\ell_2 \hf} (u^2-v^2) c_{\zeta_A} + \delta_{\ell_1 1} \delta_{\ell_2 0}\, uv\, c_{\zeta_S}.
\end{align*}
\end{theorem}

We give the grading of the non-trivial central elements:
\begin{equation}
  \begin{array}{cl}
     (0,0) & \ c \\
     (0,1) & \ c_p, \quad c_h, \quad c_{\zeta_A}, \quad c_{\zeta_S} \\
     (1,1) & \ c_x, \quad c_{\eta}, \quad c_{\kappa_A}, \quad c_{\kappa_S}
  \end{array}
  \label{center-deg}
\end{equation}
The theorem may be proved by an elementary method. 
As it is lengthy, we postpone it to the next section. 

%
\subsection{Adjoint and Superadjoint operations}

Having a class of new color superalgebras, it is important to investigate their representations. 
However, it will be beyond the scope of the present paper. 
Instead we here study some structural properties of $ \gc $ (and $\g_{\ell_1,\ell_2})$ relating to unitary representations of  \Z2-graded color superalgebra. 
Let $\g$ be a \Z2-graded color superalgebra. 
We introduce two operations in  $ \g $ which are a natural generalization of the corresponding operations in Lie superalgebras (see e.g., \cite{FrScSo} p.236). 
 
\begin{definition}{Definition}
An adjoint operation in $\g,$ denoted by  $ \dagger,  $ is a mapping $ \g \to \g $ satisfying the following conditions:
\begin{enumerate}
 \renewcommand{\labelenumi}{(\roman{enumi})}
  \item $ X_{\bm{a}}^{\dagger} \in \g_{\bm{a}} $
  \item $ (\alpha X_{\bm{a}} + \beta Y_{\bm{a}})^{\dagger} = \alpha^* X_{\bm{a}}^{\dagger} + \beta^* Y_{\bm{a}}^{\dagger}, \quad \alpha, \beta \in \mathbb{C} $  \quad ($*$ denotes the complex conjugation) 
  \item $ \llbracket X_{\bm{a}}, X_{\bm{b}} \rrbracket^{\dagger} = \llbracket X_{\bm{b}}^{\dagger}, X_{\bm{a}}^{\dagger} \rrbracket $
  \item $ (X_{\bm{a}}^{\dagger})^{\dagger} = X_{\bm{a}} $ 
\end{enumerate}
\end{definition}

\begin{definition}{Definition}
A superadjoint operation in $\g,$ denoted by $ \ddagger, $ is a mapping $ \g \to \g $ satisfying the following conditions:
  \begin{enumerate}
    \renewcommand{\labelenumi}{(\roman{enumi})}
     \item $ X_{\bm{a}}^{\ddagger} \in \g_{\bm{a}} $
     \item $ (\alpha X_{\bm{a}} + \beta Y_{\bm{b}})^{\ddagger} = \alpha^* X_{\bm{a}}^{\ddagger} + \beta^* Y_{\bm{b}}^{\ddagger} $ 
     \item $ \llbracket X_{\bm{a}}, X_{\bm{b}} \rrbracket^{\ddagger} = (-1)^{\bm{a}\cdot\bm{b}} 
     \llbracket X_{\bm{b}}^{\ddagger}, X_{\bm{a}}^{\ddagger} \rrbracket $
     \item $ (X_{\bm{a}}^{\ddagger})^{\ddagger} = (-1)^{\bm{a}\cdot\bm{a}} X_{\bm{a}} $ 
  \end{enumerate}
\end{definition}

We demonstrate the existence of the adjoint and superadjoint operations in $\gc$ (this shows the existence in $\g_{\ell_1,\ell_2},$ too) by giving an explicit example of those operations. 

\begin{lemma}{Proposition} \label{PROP:Adj}
 The following operation is an adjoint operation in the color superalgebra $ \gc: $ 
 \begin{alignat*}{3}
   L_m^{\dagger} &= L_{-m}, &\qquad\quad  P_r^{\dagger} &= P_{-r}, &\qquad\quad  X_u^{\dagger} &= X_{-u},
   \nn \\
   P_{rs}^{\dagger} &= P_{-r,-s}, & X_{uv}^{\dagger} &= -X_{-u,-v}, & T_{ru}^{\dagger} &= T_{-r,-u},
   \nn \\
   c^{\dagger} &= c,  &  c_h^{\dagger} &= c_h, & c_{\eta}^{\dagger} &= -c_{\eta}, 
 \end{alignat*}   
 \begin{alignat}{2}   
    c_p^{\dagger} &= \begin{cases}
                        -c_p & \ell_1 = 0,\\
                         c_p  & \ell_1 = 1,
                     \end{cases}
   & \qquad\qquad
    c_x^{\dagger} &= \begin{cases}
                         -c_x & \ell_2 = 0, \\
                          c_x & \ell_2 = 1,
                    \end{cases}
   \nn \\[5pt]
   c_{\kappa_A}^{\dagger} &= \begin{cases}
                                c_{\kappa_A} & \ell_1 = \ell_2 = 0, \\
                               -c_{\kappa_A} & \ell_1 =\frac{1}{2},\ \ell_2 = 0,
                             \end{cases}
   &
   c_{\kappa_S}^{\dagger} &= \begin{cases}
                                c_{\kappa_S} & \ell_1 = \ell_2 = 0, \\
                               -c_{\kappa_S} & \ell_1 =0,\ \ell_2 = 1,                                   
                             \end{cases}
   \nn \\[5pt]
   c_{\zeta_A}^{\dagger} &= \begin{cases}
                               -c_{\zeta_A} & \ell_1 = \ell_2 = 0, \\
                                c_{\zeta_A} & \ell_1 =0,\ \ell_2 = \frac{1}{2},
                             \end{cases}
   &
   c_{\zeta_S}^{\dagger} &= \begin{cases}
                               -c_{\zeta_S} & \ell_1 = \ell_2 = 0, \\
                                c_{\zeta_S} & \ell_1 =1,\ \ell_2 = 0.                                   
                             \end{cases}   
    \label{eg-adj}
 \end{alignat}
\end{lemma}

\begin{lemma}{Proposition} \label{PROP:SAdj}
 The following operation is an superadjoint operation in the color superalgebra $ \gc $ for 
 $\ell_1 \in \mathbb{N}+\hf$ and $ \ell_2 \in \mathbb{N}:$
  \begin{alignat}{3}
     L_m^{\ddagger} &= (-1)^m L_{-m}, & \qquad P_r^{\ddagger} &= (-1)^{\ell_1+r} P_{-r}, & \qquad
     X_u^{\ddagger} &= (-1)^u X_{-u}, 
     \nn \\
     P_{rs}^{\ddagger} &= (-1)^{r+s} P_{-r,-s}, & 
     X_{uv}^{\ddagger} &= (-1)^{u+v+1} X_{-u,-v}, &
     T_{ru}^{\ddagger} &= (-1)^{\ell_1+r+u+1} T_{-r,-u},
     \nn \\
     c^{\ddagger} &= c, & c_{\eta}^{\ddagger} &= -c_{\eta}, & c_{\kappa_A}^{\ddagger} &= -c_{\kappa_A},
     \nn \\
     c_x^{\ddagger} &= 
       \begin{cases}
            -c_x & \ell_2 = 0, \\
            c_x  & \ell_2 = 1.
       \end{cases}
    \label{eg-sadj}
  \end{alignat}
\end{lemma}

Proof of these propositions is straightforward. One may easily verify that the operations \eqref{eg-adj} and \eqref{eg-sadj} satisfy the definitions. 
We remark that the adjoint and superadjoint operations in $\gc$ are not unique. 
The ones in Propositions \ref{PROP:Adj} and \ref{PROP:SAdj} are understood as one of the
possibilities. 
However, it should be emphasized that the propositions show that the color superalgebra $\gc$ can have unitary representations. 
Thus one may expect that $ \gc$ can be discussed in the context of quantum physics.

%
%
%
%
\section{Proof of Theorem \ref{Thm}} \label{SEC:Proof}

By definition, central elements may appear in all the possible general Lie bracket of $ \gc.$ 
More precisely, one may consider the following relations: 
\begin{align*}
   [L_m, P_r] &= (m\ell_1-r) P_{m+r} + p(m,r), \nn \\
   [L_m, X_u] &= (m\ell_2-u) X_{m+u} + x(m,u), \nn \\
  [L_m, P_{rs}] &= (m\ell_1-r) P_{m+r,s} + (m\ell_1-s) P_{r,m+s} + p(m;r,s),  \nn \\
  [L_m, X_{uv}] &= (m\ell_2-u) X_{m+u,v} + (m\ell_2-v) X_{u,m+v} + x(m;u,v),  \nn \\
  [L_m, T_{ru}] &= (m\ell_1-r) T_{m+r,u} + (m\ell_2-u) T_{r,m+u} + t(m;r,u),  \nn \\
  \{ P_r, P_s \} &= P_{rs} + \alpha(r,s), \quad 
  [X_u, X_v] = X_{uv} + \beta(u,v), \\
  \{P_r, X_u\} &= T_{ru} + \gamma(r,u),  \\
  [P_{rs}, P_{r's'}] &= p(r,s;r',s'), \quad \;
  [X_{uv}, X_{u'v'}] = x(u,v;u',v'), \\
  [P_{rs}, X_{uv}] &= q(r,s;u,v), \quad \;\; 
  [P_{rs}, P_t] = h(r,s;t), \\
  [P_{rs}, X_u] &= \rho(r,s;u), \qquad \ \,
  [X_{uv}, P_r] = q(u,v;r), \\
  [X_{uv}, X_w] &= \eta(u,v;w), \\ 
  [P_{rs}, T_{tu}] &= \nu(r,s;t,u), \quad \ \,
  [X_{uv}, T_{rw}] = \tau(u,v;r,w), \\
  \{ T_{ru}, T_{sv} \} &= t(r,u;s,v), \quad \ \,
  [T_{ru}, P_s] = \kappa(r,u;s), \\
  \{T_{ru}, X_{v} \} &= \zeta(r,u;v).
\end{align*}
In order to have a color superalgebra, these relations must be compatible with the graded Jacobi identities. 
Enforcing this compatibility imposes constraints on the central elements. 
Below we list the constraints together with the three elements of $ \gc $ used to compute the graded Jacobi identity.  
The constraint relations are classified into five categories:

\medskip\noindent
Category I: 
\begin{itemize}
  \item $ P_{rs},\; P_{r'},\; P_{s'}:$
  \begin{equation}
    p(r,s;r',s') = 0.
  \end{equation}
\item  $ X_{uv},\; X_{u'},\;  X_{v'}:$ 
\begin{equation}
  x(u,v;u',v') = 0.
\end{equation}
\item  $ X_{uv},\;  P_{r},\;  P_{s}:$ 
\begin{equation}
  q(u,v;r,s) = 0.
\end{equation}
\item  $ P_{rs},\;  P_{t},\;  X_{u}:$ 
\begin{equation}
  \nu(r,s;t,u) = 0.
\end{equation}
\item  $ X_{uv},\;  P_{r},\;  X_{w}:$ 
\begin{equation}
  \tau(u,v;r,w) = 0.
\end{equation}
\item $ T_{ru},\;  P_{s}, \;  X_{v}:$ 
\begin{equation}
  t(r,u;s,v) = 0.
\end{equation}
\end{itemize}

\ \\

Category II:
\begin{itemize}
\item  $ L_m,\;  L_n,\;  P_r:$ 
\begin{equation}
   (n\ell_1 -r)\, p(m,n+r) - (m\ell_1 -r)\, p(n,m+r) -(m-n)\, p(m+n,r) = 0. 
   \label{J-LLP}
\end{equation}
\item   $L_m,\; L_n,\;  X_u:$
\begin{equation}
   (n\ell_2 -u)\, x(m,n+u) - (m\ell_2 -u)\, x(n,m+u) -(m-n)\, x(m+n,u) = 0. 
   \label{J-LLX}
\end{equation}
\end{itemize}
Category III:
\begin{itemize}
\item  $ P_r,\;  P_s,\;  P_t:$ 
\begin{equation}
  h(r,s;t) + h(s,t;r) + h(t,r;s) = 0. \label{J-PPP}
\end{equation}
\item  $ X_u,\;  X_v,\;  X_w:$ 
\begin{equation}
  \eta(u,v;w) + \eta(v,w;u) + \eta(w,u;v) = 0.  \label{J-XXX}
\end{equation}
\item $ L_m,\;  P_{rs},\;  P_t:$ 
\begin{align}
  (m\ell_1-r)\, h(m+r,s;t) &+ (m\ell_1-s)\, h(r,m+s;t) \nn \\
  & + (m\ell_1-t)\, h(r,s;m+t) = 0. 
  \label{J-LP2P}
\end{align}
\item $ L_m,\;  X_{uv},\;  X_w:$ 
\begin{align}
  (m\ell_2-u)\, \eta(m+u,v;w) &+ (m\ell_2-v)\, \eta(u,m+v;w) \nn \\
  & + (m\ell_2-w)\, \eta(u,v;m+w) = 0. 
  \label{J-LX2X}
\end{align}
\end{itemize}
Category IV:
\begin{itemize}
\item $ P_r,\;  P_s,\;  X_u:$ 
\begin{equation}
  \kappa(r,u;s) + \kappa(s,u;r) + \rho(r,s;u) = 0. \label{J-PPX}
\end{equation}
\item $ P_r,\;  X_u,\;  X_v:$ 
\begin{equation}
  \zeta(r,u;v)-\zeta(r,v;u) + q(u,v;r) = 0. \label{J-PXX}
\end{equation}
\item $ L_m,\;  P_{rs},\;  X_u:$ 
\begin{align}
  (m\ell_2-u)\, \rho(r,s;m+u) &+ (m\ell_1-r)\, \rho(m+r,s;u) \nn \\ & + (m\ell_1-s)\, \rho(r,m+s;u) = 0. 
  \label{J-LP2X}
\end{align}
\item  $ L_m,\;  X_{uv},\;  P_r:$ 
\begin{align}
  (m\ell_2-u)\, q(m+u,v;r) &+ (m\ell_2-v)\, q(u,m+v;r) \nn \\
  & +(m\ell_1-r)\, q(u,v;m+r) = 0
  \label{J-LX2P}
\end{align}
\item $ L_m,\;  T_{ru},\;  P_s:$ 
\begin{align}
  (m\ell_2 -u)\, \kappa(r,m+u;s) &+ (m\ell_1-r)\, \kappa(m+r,u;s)   \nn \\
  & + (m\ell_1-s)\, \kappa(r,u;m+s) = 0. \label{J-LT2P}
\end{align}
\item  $ L_m,\;  T_{ru},\;  X_v:$ 
\begin{align}
   (m\ell_2-u)\, \zeta(r,m+u;v)   & + (m\ell_2-v)\, \zeta(r,u;m+v) 
   \nn \\   
   & + (m\ell_1-r)\, \zeta(m+r,u;v) = 0.
   \label{J-LT2X}
\end{align}
\end{itemize}
Category V:
\begin{itemize}
\item $ L_m,\;  P_r,\;  P_s:$ 
\begin{equation}
  p(m;r,s) - (m\ell_1-r)\, \alpha(m+r,s) - (m\ell_1-s)\, \alpha(r,m+s) = 0. 
  \label{J-LPP}
\end{equation}
\item $ L_m,\;  X_u,\;  X_v:$ 
\begin{equation}
  x(m;u,v)-(m\ell_2-u)\, \beta(m+u,v) - (m\ell_2-v)\, \beta(u,m+v) = 0.
  \label{J-LXX}
\end{equation}
\item  $ L_m,\;  P_r,\;  X_u:$ 
\begin{equation}
  t(m;r,u) - (m\ell_1-r)\, \gamma(m+r,u) - (m\ell_2-u)\, \gamma(r,m+u) = 0.
  \label{J-LPX}
\end{equation}
\item  $ L_m,\;  L_n,\;  P_{rs}:$ 
\begin{align}
  & (n\ell_1-r)\, p(m;n+r,s) + (n\ell_1-s)\, p(m;r,n+s)  - (m\ell_1-r)\, p(n;m+r,s) \nn \\
  & - (m\ell_1-s)\, p(n;r,m+s) - (m-n)\, p(m+n;r,s) = 0. 
  \label{J-LLP2}
\end{align}
\item  $ L_m,\;  L_n,\;  X_{uv}:$ 
\begin{align}
  & (n\ell_2-u)\, x(m;n+u,v) + (n\ell_2-v)\, x(m;u,n+v)  - (m\ell_2-u)\, x(n;m+u,v) \nn \\
  &  - (m\ell_2-v)\, x(n;u,m+v)  - (m-n)\, x(m+n;u,v) = 0.
  \label{J-LLX2}
\end{align}
\item $ L_m,\;  L_n,\;  T_{ru}:$ 
\begin{align}
  & (n\ell_1-r) \,t(m;n+r,u) + (n\ell_2-u)\, t(m;r,n+u) - (m\ell_1-r)\, t(n;m+r,u)\nn \\
  &  - (m\ell_2-u)\, t(n;r,m+u)  - (m-n)\, t(m+n;r,u) = 0. 
  \label{J-LLT2}
\end{align}
\end{itemize}
The graded Jacobi identities for other combinations of elements of $\gc $ do not produce any more
new relations.
Thus solving the relations \eqref{J-LLP} to \eqref{J-LLT2} gives the complete classification of the
central extensions of $\mathfrak{g}_{\ell_1,\ell_2}.$
We prove the theorem by adopting this approach in what follows.  

The relations in Category I show that the extensions are trivial. 
In fact, the relations in Category V also show the same.  

\medskip\noindent
(i) $ \alpha(r,s),\; \beta(u,v),\; \gamma(r,u),\; p(m;r,s),\; x(m;u,v),\; t(m;r,u) $ are trivial (Category V). 

These extensions are absorbed by redefining $ P_{rs},\; X_{uv} $ and $ T_{ru} $ as follows:
\begin{equation}
   P'_{rs} = P_{rs} + \alpha(r,s), \qquad 
   X'_{uv} = X_{uv} + \beta(u,v), \qquad 
   T'_{ru} = T_{ru} + \gamma(r,u).
\end{equation}
The absorption of $ p(m;r,s) $ is verified by computing $ [L_m, P'_{rs}] $ with \eqref{J-LPP}. 
The relation \eqref{J-LLP2} is then trivial if all $p$'s are removed using \eqref{J-LPP}.  
The same argument is also applied for $ x(m;u,v)$ and $ t(m;r,u)$ for the analogous equations
\eqref{J-LXX}, \eqref{J-LPX}, \eqref{J-LLX2} and \eqref{J-LLT2}.  

\medskip\noindent
(ii) $ p(m,r) $ and $ x(m,u) $ (Category II).  

Set $m=0 $ in \eqref{J-LLP} gives, if $n+r \neq 0$,
\[
  p(n,r) = -\frac{n \ell_1 - r}{n+r} p(0,n+r).
\]
This extension is trivial since it is absorbed by the redefinition $ P'_r = P_r + r^{-1} p(0,r) $ where $ r \neq 0. $ Therefore one may write
\begin{equation}
  p(n,r) = \delta_{n+r,0}\, p(n).
\end{equation} 
This implies that $r \in \mathbb{Z} $ so that $ \ell_1 $ is a non-negative integer. 
Equation \eqref{J-LLP} then becomes
\begin{equation}
   (n\ell_1+m+n) p(m) - (m\ell_1+m+n) p(n) -(m-n) p(m+n) = 0.
   \label{p-recrel}
\end{equation}
Setting $m=0$ in \eqref{p-recrel}, we see that $ p(0) = 0 $ for any $ \ell_1. $ 
We also see from \eqref{p-recrel} by setting $m+n=0$ that 
$
  m \ell_1 (p(m)+p(-m)) = 0.
$ We thus have
\begin{equation}
   p(-m) = -p(m) \quad \text{if} \quad \ell_1 \neq 0 \label{pmpm}
\end{equation}
Now we set $m=1$ then \eqref{p-recrel} becomes
\begin{equation}
    (n-1) p(n+1) - (\ell_1 +n+1) p(n) + (n\ell_1+n+1) p(1) = 0. \label{p-eq1}
\end{equation}
In \eqref{p-recrel}, set $ m = -1 $ and replace $n$ with $n+1$ to obtain the relation
\begin{equation}
  (\ell_1-n) p(n+1) + (n+2) p(n) + (n\ell_1 + \ell_1 + n) p(-1) = 0.
  \label{p-eq2}
\end{equation}
Two relations \eqref{p-eq1} and \eqref{p-eq2} give the identity
\begin{align}
  (\ell_1+2) (\ell_1-1) p(n) &= (\ell_1 - n)(\ell_1 n + n+1) p(1) - ((\ell_1+1)n^2 - \ell_1 - n) p(-1)
  \label{p-eq3} \\
  & = (\ell_1+2) (\ell_1-1)n p(1) \nn 
\end{align}
where \eqref{pmpm} was used in the second equality. It follows that
\begin{equation}
    p(n) = n p(1) \quad \text{if} \quad \ell_1 \neq 0, 1.
\label{linp}
\end{equation}
However, this extension is trivial since the redefinition $ P'_0 =  P_0 + (\ell_1+1)^{-1} p(1) $
absorbs the $p(n)$ being linear in $n$. Therefore we conclude that there exists no non-trivial
extension involving $p(n)$ for $\ell_1 \neq 0, 1. $ 

For $ \ell_1 = 0, $ \eqref{pmpm} does not hold true but \eqref{p-eq3} with $\ell_1 = 0 $ is still valid and becomes
\begin{equation}
    p(n) = \hf n^2 (p(1) + p(-1)) + \hf n (p(1)-p(-1)).
\end{equation} 
It is easy to verify that this $p(n)$ satisfies \eqref{p-recrel} with $\ell_1 = 0.$  
The second term (linear in $n$) of $p(n)$ is a trivial extension and $ \ell_1 = 0 $ implies $ p(1) +
p(-1) \neq 0 $ unless $p(1) = p(-1) = 0. $ Therefore there exists a non-trivial extension if $ \ell_1 = 0. $ 

If we set $\ell_1 = 1,$ then \eqref{p-recrel} becomes the same relation as the one for the central
extension of the Virasoro algebra.
Thus there exists a non-trivial extension for $ \ell_1 = 1.$ 

One may repeat the same argument for $ x(m,u) $ since $ x(m,u) $ satisfies the same relation as
$p(m,r)$ provided that $ \ell_1 $ is replaced with $ \ell_2.$

\medskip\noindent
(iii) $ q(u,v;r)$ and $ \zeta(r,u;v) $ (Category IV). 
 
 As seen from \eqref{J-LX2P} and \eqref{J-LT2X}, $ q(u,v;r)$ and $ \zeta(r,u;v) $ satisfy the same
relation. The only difference is the symmetric property, namely $ q(u,v;r) $ is antisymmetric in the
first two arguments while $ \zeta(r,u;v) $ does not have a particular symmetry.
Set $ m = 0, $ then \eqref{J-LX2P} and \eqref{J-LT2X} become
\begin{equation}
   (u+v+r) q(u,v;r) = (r+u+v) \zeta(r,u;v) = 0.
\end{equation}
Taking into account the antisymmetric property of $ q(u,v;r)$ one may write
\begin{align*}
   q(u,v;r) &= \delta_{u+v+r,0}\, q(u,v), \qquad  q(u,v) = -q(v,u),
   \\
   \zeta(r,u;v) &= \delta_{r+u+v,0}\, \zeta(u,v).
\end{align*}
This implies $ r \in \mathbb{Z} $ so that $ \ell_1 $ is a non-negative integer. 
With this observation the equations \eqref{J-PXX}, \eqref{J-LX2P} and \eqref{J-LT2X} read
\begin{align}
    & \zeta(u,v) - \zeta(v,u) + q(u,v) = 0, \label{zetaq}
    \\
    & (m\ell_2-u) q(m+u,v)+(m\ell_2-v) q(u,m+v) + (m\ell_1 + m+u+v) q(u,v) = 0, \label{q-rel}
    \\
    & (m\ell_2-u) \zeta(m+u,v)+(m\ell_2-v) \zeta(u,m+v) + (m\ell_1 + m+u+v) \zeta(u,v) = 0.    
    \label{zeta-rel}
\end{align}

$\zeta(u,v)$ may be decomposed into the symmetric part $(\zeta_S)$ and the antisymmetric part $(\zeta_A)$:
\begin{equation}
  \zeta(u,v) = \zeta_S(u,v) + \zeta_A(u,v).
\end{equation}
Substituting this into \eqref{zetaq} one may express $ q(u,v) $ in terms of $\zeta_A(u,v):$
\begin{equation}
   q(u,v) + 2\zeta_A(u,v) = 0.
\end{equation}
It follows that $ \zeta_A(u,v)$ and $ \zeta_S(u,v) $ satisfy \eqref{zeta-rel} separately.  
Therefore all possible central extensions are obtained by solving \eqref{zeta-rel} for $ \zeta_A $ and $\zeta_S. $ 

\medskip\noindent
(iii-a) Antisymmetric $ \zeta_A(u,v). $

We first solve \eqref{zeta-rel} for $ \zeta_A. $ 
Set $ m = 1 $ and $ v=\ell_2,$ so that \eqref{zeta-rel} becomes
\begin{equation}
   (\ell_2-u)\, \zeta_A(u+1,\ell_2) + (\ell_1+\ell_2+1+u)\, \zeta_A(u,\ell_2) = 0.
   \label{m1zeta}
\end{equation}
Varying $u$ from $ \ell_2 $ to $ -\ell_1 - \ell_2 -1 $ and $ \zeta_A(\ell_2,\ell_2) = 0 $ one sees from \eqref{m1zeta} that
\begin{equation}
   \zeta_A(u,\ell_2) = 0, \quad -\ell_1-\ell_2 \leq u \leq \ell_2  \label{zeta0-1} 
\end{equation}
One may also determine that $ \zeta_A(u,\ell_2) = 0 $ beyond the range of $u$ given in
\eqref{zeta0-1} as follows,
except the two cases $ (\ell_1,\ell_2) = (0,0), (0,\hf).$
For $ u \geq  \ell_2+1 $ \eqref{m1zeta} is solved to give
\begin{equation}
   \zeta_A(u,\ell_2) = \frac{(u+\ell_1+\ell_2)!}{(\bar{\ell}+1)! (u-\ell_2-1) !}\, \zeta_A(\ell_2+1,\ell_2), 
   \qquad \bar{\ell} = \ell_1 + 2\ell_2
   \label{zeta-sol1}
\end{equation}
while for $ u \leq -\ell_1 - \ell_2 - 1 $
\begin{equation}
   \zeta_A(-u, \ell_2) = \frac{(u+\ell_2)!}{(\bar{\ell}+1)! (u-\ell_1-\ell_2-1)!}\, \zeta_A(-\ell_1 - \ell_2 - 1,\ell_2), \qquad 
   u \geq \ell_1+\ell_2+1.
   \label{zeta-sol22}
\end{equation}
Set $(m,u,v) = (2,\ell_2+2,\ell_2), $ then \eqref{zeta-rel} becomes
\[
  (\ell_2-2) \zeta_A(\ell_2+4,\ell_2) + 2(\ell_1 + \ell_2+2) \zeta_A(\ell_2+2,\ell_2) = 0
\]
and by \eqref{zeta-sol1} this gives
\[
   (\bar{\ell}+2)\, \bar{\ell} \,( \ell_1 (\ell_2-2) + (2\ell_2-1)(\ell_2+2)) \zeta_A(\ell_2+1,\ell_2) = 0.
\]
It follows that 
\[
    \zeta_A(\ell_2+1,\ell_2) = 0, \qquad (\ell_1,\ell_2) \neq (0,0), \; \Big(0,\hf \Big)
\]
and $ \zeta_A(u, \ell_2) = 0 $ for $ u > \ell_2. $ 

Now we set $(m,u,v) = (-\bar{\ell}, -\ell_1-\ell_2,\ell_2),$ then \eqref{zeta-rel} becomes with the aide of \eqref{zeta-sol22}
\begin{align*}
  & ((1-\bar{\ell}) \ell_2 + \ell_1) \zeta_A(-2\ell_1-3\ell_2,\ell_2)
  \\
  & \qquad 
   =  ((1-\bar{\ell}) \ell_2 + \ell_1) \frac{(2\bar{\ell})!}{(\bar{\ell}-1)! (\bar{\ell}+1)!} \,
    \zeta_A(-\ell_1-\ell_2-1,\ell_2) = 0.
\end{align*}
It follows that 
\[
   \zeta_A(-\ell_1-\ell_2-1,\ell_2) = 0, \qquad (\ell_1,\ell_2) \neq (0,0), \; \Big(0,\hf \Big)
\]
and $ \zeta_A(u, \ell_2) = 0 $ for $ u < -\ell_1-\ell_2. $ 
Therefore we have proved that
\begin{equation}
  \zeta_A(u,\ell_2) = 0, \quad \text{for}\ \forall u,  
  \quad (\ell_1,\ell_2) \neq (0,0),\ \Big(0,\hf \Big).  \label{zeta0-2}
\end{equation}

  We now set $ v = \ell_2, $  then \eqref{zeta-rel} gives
\[
  (m\ell_2-u) \zeta_A(m+u,\ell_2) + \ell_2(m-1) \zeta_A(u,m+\ell_2) + (m(\ell_1+1) + \ell_2 + u) \zeta_A(u,\ell_2) = 0. 
\]
The first and the last term of this relation vanish due to \eqref{zeta0-2} so that we have
$ \zeta_A(u,m+\ell_2) = 0 $ for $ m \neq 1. $ Since $u$ is arbitrary one may set $u=\ell_2+1$, then
$\zeta_A(\ell_2+1, m+\ell_2) = 0 $ which corresponds to the $m =1$ case. 
Thus we have shown 
\begin{equation}
  \zeta_A(u,v) = 0, \quad \text{for}\ \forall u, v,  
  \quad (\ell_1,\ell_2) \neq (0,0),\ \Big(0,\hf \Big).  \label{zeta0-3}
\end{equation}

We proceed to investigate the two exceptional cases of $ (\ell_1, \ell_2). $ 
We start with $ (\ell_1,\ell_2) = (0,\hf). $ 
For these values of $ (\ell_1, \ell_2),$ equation \eqref{m1zeta} becomes
\begin{equation}
  \Big( \hf -u \Big) \zeta_A \Big(u+1,\hf\Big)+ \Big(\frac{3}{2} + u \Big) \zeta_A \Big(u,\hf\Big) = 0.
  \label{0-half-rel}
\end{equation}
Setting $u=-\hf$ it is immediate to see from this relation that
\begin{equation}
  \zeta_A \Big(-\hf,\hf \Big) = \zeta_A\Big(\hf,\hf \Big) = 0. 
\end{equation}
The relation \eqref{0-half-rel} is easily solved, for $ u > \hf $ and $ u < -\hf $,
to give
\begin{equation}
   \zeta_A \Big(u,\hf\Big) = \hf \Big( u^2 - \frac{1}{4} \Big) \zeta_A \Big(\frac{3}{2}, 
    \hf\Big),
   \quad
   \zeta_A \Big(-u,\hf\Big) = \hf \Big( u^2 - \frac{1}{4} \Big) \zeta_A \Big(-\frac{3}{2}, 
    \hf\Big)
    \label{zeta-sol}
\end{equation}
where $ u > \frac{3}{2}. $ Note that these relations also hold true for $ u=\hf. $   
We now show that
\begin{equation}
  \zeta_A \Big(\frac{3}{2}, \hf\Big) = \zeta_A \Big(-\frac{3}{2}, \hf\Big). 
  \label{zeta-ini2}
\end{equation}
Set $ (m,u,v) = (3,-\frac{3}{2}, \hf), \ (-3,-\frac{3}{2}, \frac{7}{2}), $ 
then from \eqref{zeta-rel} we obtain the following relations:
\begin{align*}
   & 3 \zeta_A\Big( \frac{3}{2},\hf \Big) + \zeta_A\Big(-\frac{3}{2}, \frac{7}{2} \Big) + 2 \zeta_A\Big(-\frac{3}{2},\hf \Big) = 0,
   \\
   & 5 \zeta_A\Big(-\frac{3}{2},\hf \Big) + \zeta_A\Big(-\frac{3}{2},\frac{7}{2} \Big) = 0.
\end{align*}
Equation \eqref{zeta-ini2} follows immediately from these relations. 
Combining \eqref{zeta-sol} and \eqref{zeta-ini2} we have
\begin{equation}
   \zeta_A \Big(u,\hf \Big) = \hf \Big( u^2 - \frac{1}{4} \Big) \zeta_A\Big(\frac{3}{2}, \hf\Big),
\quad \text{for} \ \forall u. 
   \label{zeta-sol2}
\end{equation}
Now set $ v = \hf$ in \eqref{zeta-rel} to give
\[
  \Big( \frac{m}{2} -u \Big) \zeta_A \Big( m+u,\hf \Big) + \hf (m-1) \zeta_A \Big(u,m+\hf \Big) + \Big( m+u+\hf \Big) \zeta_A\Big(u,\hf\Big) = 0.
\]
This relation, together with \eqref{zeta-sol2}, shows that if $ m \neq 1,$ then 
\[
  \zeta_A \Big(u,m+\hf\Big) = \hf \Big( u^2 - \big(m+\hf\big)^2 \Big) \zeta_A\Big( \frac{3}{2},\hf
\Big), \quad \forall u.
\]
Since $ u$ is arbitrary one may set $ u = \frac{3}{2} $ and see that the above relation also holds true for $m=1$. 
Thus we have shown that for $ (\ell_1,\ell_2) = (0,\hf) $
\begin{equation}
    \zeta_A(u,v) =  (u^2-v^2) c_{\zeta_A}, \quad \forall u, v.
\end{equation}
Next we study the case of $ (\ell_1,\ell_2) = (0,0). $ 
Note that $ u, v \in \mathbb{Z} $ in this case. 
Equation \eqref{m1zeta} becomes
\[
u \zeta_A(u+1,0) = (u+1) \zeta_A(u,0)
\]
and this relation is easily solved: $ \zeta_A(u,0) = \alpha u $ with $ \alpha = \zeta_A(1,0) $ for any $u.$  
Setting $ m=-1, v=1 $ in \eqref{zeta-rel} gives
\[
  u \zeta_A(u-1,1) + \zeta_A(u,0) - u \zeta_A(u,1) = 0.
\]
It follows that if $ u \neq 0$
\begin{equation}
    \zeta_A(u,1) - \zeta_A(u-1,1) = \zeta_A(1,0).
\end{equation}
One may solve this for positive and negative $u$ as follows:
\begin{equation}
  \zeta_A(u,1) = (u-1) \zeta_A(1,0), \quad
    \zeta_A(-u,1) = -(u-1) \zeta_A(1,0) + \zeta_A(-1,1), 
    \quad u \geq 1. 
    \label{zeta0-4}
\end{equation}
By setting $(m,u,v) = (-2,-1,1)$, we see from \eqref{zeta-rel} that $ \zeta_A(-1,1) = -2\zeta_A(1,0).
$ Together with \eqref{zeta0-4}, we obtain that
\begin{equation}
    \zeta_A(u,1) = (u-1) \zeta_A(1,0), \quad \forall u.  \label{zeta0-5}
\end{equation}
Finally, using \eqref{zeta0-5}, we set $ v = 1 $ in \eqref{zeta-rel} to conclude that
\begin{equation}
\zeta_A(u,v) = (u-v) \zeta_A(1,0), \quad \forall u, v.
\end{equation}
Therefore we obtain a non-trivial central extension for $ \ell_1 = \ell_2 = 0.  $

\medskip
In summary, there exist two central extensions for $ \zeta_A $ and $ q$:
\[
 \begin{array}{c|c}
    (\ell_1,\ell_2) & \zeta_A(u,v) \\ \hline
    (0,0)  & (u-v) c_{\zeta_A} \\
    (0,\hf)  & (u^2-v^2) c_{\zeta_A}
 \end{array}, 
\]
with $ q(u,v) = -2\zeta_A(u,v)$.

\medskip\noindent
(iii-b) Symmetric $ \zeta_S(u,v). $

Set $ u=v= \ell_2,$ then \eqref{zeta-rel} becomes
\begin{equation}
   2 \ell_2 (m-1)\zeta_S(\ell_2 +m,\ell_2 ) + (m\ell_1+m+2 \ell_2) \zeta_S(\ell_2,\ell_2) = 0. 
   \label{z-sym-eq1}
\end{equation} 
Further setting $m=1$ gives the simple relation $ (\bar{\ell}+1) \zeta_S(\ell_2,\ell_2 ) = 0$, which
shows that
\begin{equation}
   \zeta_S(\ell_2,\ell_2) = 0, \qquad \forall \ell_1, \ell_2.  \label{z-sym-eq2}
\end{equation} 
Returning to \eqref{z-sym-eq1}, we see that $ \zeta_S(\ell_2+m,\ell_2) = 0 $ if $ \ell_2 \neq 0 $ and $ m \neq 1. $ 
This is also true for $m=1$ since setting $(m,u,v) = (1, \ell_2+1,\ell_2)$ in \eqref{zeta-rel} gives
$ (\bar{\ell}+2) \zeta_S(\ell_2 +1,\ell_2) = 0$.
We thus have obtained
\begin{equation}
    \zeta_S(\ell_2 +m,\ell_2) = 0, \quad \forall m, \ell_1 \quad \text{if} \quad \ell_2 \neq 0.
    \label{z-sym-eq3}
\end{equation}
Now we set $ v = \ell_2, $ then we have from \eqref{zeta-rel} and \eqref{z-sym-eq3} that
$ \ell_2  (m-1) \zeta_S(u,\ell_2 +m) = 0.$ 
It follows that if $ \ell_2 \neq 0 $ and $ m \neq 1$ then  $ \zeta_S(u,\ell_2 +m) = 0 $ for any $u, \ell_1.$ 
This is also true for $ m=1,$ which is seen by noting that the relation is true for  $ u = \ell_2 +1
$ and exchanging the two arguments which gives   $ \zeta_S(\ell_2 +m,\ell_2 +1) =0.$  Therefore we
have proved that
\begin{equation}
  \zeta_S(u,v) = 0, \quad \forall u, v, \ell_1 \quad \text{if} \quad \ell_2 \neq 0.
\end{equation}
Namely, there exist no non-trivial central extension $\zeta_S$ if $ \ell_2 \neq 0. $ 

Next we investigate the case of $ \ell_2 = 0. $ 
Note that $ u, v \in \mathbb{Z} $ in this case. 
Set $ m+u=v=0, $ then \eqref{zeta-rel} and \eqref{z-sym-eq2} give $ u\ell_1 \zeta_S(u,0) = 0 $ for any $u$ and $\ell_1. $ 
Thus we have
\begin{equation}
   \zeta_S(u,0) = 0, \quad \forall u \quad \text{if} \quad \ell_1 \neq 0. \label{z-sym-eq4}
\end{equation} 
Set $u=v,$ then \eqref{zeta-rel} becomes
\begin{equation}
      (m\ell_1+m+2u) \zeta_S(u,u) - 2u\zeta_S(u+m,u) = 0.  \label{z-sym-eq5}
\end{equation}
By setting $ m+u=0$ in this relation we obtain  that $ u(\ell_1-1) \zeta_S(u,u) = 0 $ for any $u.$  
Since \eqref{z-sym-eq4} was  used to obtain this, we have shown that
\begin{equation}
    \zeta_S(u,u) = 0, \quad \forall u \quad \text{if} \quad \ell_1 \neq 0, 1.
\end{equation}
Returning to \eqref{z-sym-eq5} we immediately conclude that
\begin{equation}
     \zeta_S(u,v) = 0, \quad \forall u, v \quad \text{if} \quad \ell_1 \neq 0, 1.
\end{equation}
Therefore the possibility of non-trivial extensions is restricted to $ \ell_1 = 0 $ or 1.

Let us study the case of $ \ell_1 = 1. $ 
Set $v=1,$ then \eqref{zeta-rel} becomes
\begin{equation}
   (2m+1+u) \zeta_S(u,1) - u \zeta_S(u+m,1) - \zeta_S(u,m+1) = 0.
   \label{z-sym-eq6}
\end{equation}
Since \eqref{z-sym-eq4} holds true for $\ell_1 =1,$  we obtain a  recurrence relation from \eqref{z-sym-eq6} by setting $ m=-1: $
\[
   (u-1) \zeta_S(u,1) - u \zeta_S(u-1,1) = 0.
\]
This is easily solved to give $ \zeta_S(u,1) = u \alpha $ with $ \alpha =\zeta_S(1,1)$ being a constant. 
Substituting the solution into \eqref{z-sym-eq6} then gives $ \zeta_S(u,m+1) = u(m+1) \alpha. $ 
We have thus proved that if $ \ell_1 = 1 $ and $ \ell_2 = 0$ there exists a non-trivial extension given by
\begin{equation}
    \zeta_S(u,v) = uv\, c_{\zeta_S}, \quad \forall u, v
\end{equation}

We turn to the case of $ \ell_1 = 0. $
Set $ v = 0, $ then \eqref{zeta-rel} has the simple form
\begin{equation}
   (m+u) \zeta_S(u,0) - u \zeta_S(m+u,0) = 0.   \label{z-sym-eq7}
\end{equation}
%
%
Setting $ u=0,$ we observe that $ \zeta_S(0,0) = 0. $ 
Having this in mind, the solution to \eqref{z-sym-eq7} is easily found: 
$ \zeta_S(u,0) = u \alpha $ with $ \alpha = \zeta_S(1,0) $ being a constant. 
Now we set $ u=v, $ then \eqref{zeta-rel} becomes
\begin{equation}
   (m+2u) \zeta_S(u,u) - 2u \zeta_S(u+m,u) = 0.  \label{z-sym-eq8}
\end{equation}
By setting $ u+m=0, $ one may see that $ \zeta_S(u,u) = 2 u \alpha $ for any $u.$ 
Substituting this back into \eqref{z-sym-eq8} we obtain $ \zeta_S(u+m,u) = (2u+m) \alpha $ for all $u, m. $ 
We have therefore found a non-trivial extension given by 
\begin{equation}
    \zeta_S(u,v) = (u+v) c_{\zeta_S}, \quad \forall u, v
\end{equation}

\medskip
In summary, there exist two central extensions $ \zeta_S: $
\[
 \begin{array}{c|c}
    (\ell_1,\ell_2) & \zeta_S(u,v) \\ \hline
    (0,0)  & (u+v) c_{\zeta_S} \\
    (1,0)  & uv c_{\zeta_S}
 \end{array}
\]

\medskip\noindent
(iv) $ \rho(r,s;v) $ and $ \kappa(r,u;s) $ (Category IV). 

Set $m=0, $ then \eqref{J-LP2X} and \eqref{J-LT2P} become
\[
    (u+r+s) \rho(r,s;u) = (r+u+s) \kappa(r,u;s) = 0.
\]
It follows that the non-trivial extensions exist only if $ r+s+u = 0 $ which means that 
$ u \in \mathbb{Z} $ and $ \ell_2 $ is a non-negative integer. 
Recalling that $\rho$ is symmetric with respect to the first two arguments, one may write
\begin{align*}
   & \rho(r,s;u) = \delta_{r+s+u,0} \rho(r,s), \qquad \rho(r,s) = \rho(s,r),
   \\
   & \kappa(r,u;s) = \delta_{r+s+u,0} \kappa(r,s).
\end{align*}
With these expressions \eqref{J-PXX}, \eqref{J-LP2X} and \eqref{J-LT2P} yield
\begin{align}
  & \kappa(r,s) + \kappa(s,r) + \rho(r,s) = 0,
  \nn \\
  & (m\ell_1 - r)\rho(m+r,s) + (m\ell_1-s) \rho(r,m+s) + (m\ell_2+m+r+s) \rho(r,s) = 0,
  \nn \\
  & (m\ell_1 - r)\kappa(m+r,s) + (m\ell_1-s) \kappa(r,m+s) + (m\ell_2+m+r+s) \kappa(r,s) = 0.  \label{kappa-rel}
\end{align}
Decomposing $ \kappa $ into the symmetric and antisymmetric parts $ \kappa = \kappa_S + \kappa_A,$ 
one may see that 
\[
    \rho(r,s) + 2 \kappa_S(r,s)  = 0
\]
and $ \kappa_S $ and $ \kappa_A $ satisfy \eqref{kappa-rel} separately. 
The relation \eqref{kappa-rel} is the same as \eqref{zeta-rel} provided that the role of $ \ell_1 $ and $ \ell_2$ is exchanged. 
Therefore one may rely on the same arguments as in (iii). We will not repeat it and present only the result. 
There exist the following central extensions:
\[
  \begin{array}{c|cc}
    (\ell_1,\ell_2) & \kappa_S(r,s) & \kappa_A(r,s) \\ \hline
      (0,0)         & (r+s) c_{\kappa_S} & (r-s) c_{\kappa_A} \\
      (\hf,0)       &   0            & (r^2-s^2) c_{\kappa_A} \\
      (0,1)         &   rs \, c_{\kappa_S}  &  0
  \end{array}
\]

\medskip\noindent
(v) $ h(r,s;t) $ and $ \eta(u,v;w) $ (Category III). 

These extensions have a definite symmetry:
\begin{equation}
  h(r,s;t) = h(s,r;t), \qquad \eta(u,v;w) = - \eta(v,u;w).
\end{equation}
Setting $m=0$, we obtain the relations from \eqref{J-LP2P} and \eqref{J-LX2X}:
\[
 (r+s+t) h(r,s;t) = (u+v+w) \eta(u,v;w) = 0.
\]
Thus the extension $h(r,s;t)$ (respectively $\eta(u,v;w)$) exists only if $\ell_1$ (respectively
$\ell_2$) is a non-negative integer.  
One may write
\begin{alignat*}{2}
    h(r,s;t) &= \delta_{r+s+t,0} \,h(r,s), & \qquad h(r,s) &= h(s,r),
   \\
    \eta(u,v;w) &= \delta_{u+v+w,0} \,\eta(u,v), & \quad \eta(u,v) &= - \eta(v,u).
\end{alignat*}
With these, equations \eqref{J-PPP}, \eqref{J-XXX}, \eqref{J-LP2P} and \eqref{J-LX2X} become
\begin{align}
  & h(r,s) + h(s,-r-s) + h(-r-s,r) = 0,
  \nn \\
  &(m\ell_1-r) h(m+r,s) + (m\ell_1-s) h(r,m+s) + (m \ell_1+m+r+s) h(r,s) = 0, \label{h-eq1}
  \\
  & \eta(u,v) + \eta(v,-u-v) + \eta(-u-v,u) = 0,
  \nn \\
  & (m\ell_2-u) \eta(m+u,v) + (m\ell_2-v) \eta(u,m+v) + (m \ell_2 +m+u+v) \eta(u,v) =0.
  \label{h-eq2}
\end{align}
The relations \eqref{h-eq1} and \eqref{h-eq2} have the same form as \eqref{zeta-rel} with $ \ell_1 = \ell_2. $ 
Thus we can apply the procedure same as (iii) and obtain the following non-trivial extensions:
\begin{equation}
  h(r,s) = \delta_{\ell_1 0} (r+s) c_h,
  \qquad
  \eta(u,v) = \delta_{\ell_2 0} (u-v) c_{\eta}.
\end{equation}
This completes the proof of the theorem.

%
%
%
%
\section{Concluding remarks} \label{SEC:CR}

In order to demonstrate the utility of color superalgebras in a mathematical setting, 
we introduced the novel class of \Z2-graded color superalgebras $\g_{\ell_1,\ell_2}$ of infinite
dimension which are realized in the UEA of the superalgebra $\mathfrak{v}_{\ell_1,\ell_2}. $ 
Then a classification of all possible central extensions of $\mathfrak{g}_{\ell_1,\ell_2} $ was given. 
We observed that non-trivial central extensions exist only when one of $\ell_1, \ell_2$ vanishes or
equals unity. 
This reflects the fact that the structure of $\mathfrak{g}_{\ell_1,\ell_2} $ is fixed strongly by the Virasoro generators $ L_m $ (see \eqref{Color-def}). 
If one of $\ell_1, \ell_2$ vanishes, then the constraint on the central elements from the graded
Jacobi identity is relaxed significantly. 
This is also the case if one of $\ell_1$ or $ \ell_2 $ is equal to unity. 
This is the reason why the non-trivial extensions exist only for small values of parameters. 

 We also showed that the algebras $\gc$ and $ \g_{\ell_1,\ell_2}$ admit adjoint and superadjoint operations.  Thus they may have unitary representations. 
As is well-known, the importance of representation theory  is not only in mathematics but also in
physics. Many physical applications of algebraic objects are made through their representations, 
especially unitary representations which are essential in quantum physics. 
The study of the representations of $\gc$ still remains an open problem, but the existence of
unitary representations allows for the possibility that $\gc$ has physical applications. 
To the best of our knowledge, connections between infinite dimensional color superalgebras and
physics have been considered only in \cite{zhe}. 
Since $\gc$ is a color extension of the Virasoro algebra, it would be possible to discuss its
connection to string theory or conformal field theory. 
To this end, we need to understand the representations of $\gc$ more deeply. 
This will be the subject of future work.

%
%
%
\section*{Acknowledgements}

N.A. and J.S. would like to thank P.S. Isaac for his warm hospitality at The University of Queensland, where
this work was started. J.S. would also like to acknowledge the support of the School of Mathematics and
Physics at The University of Queensland for the Ethel Raybould Fellowship. P.S.I. would also like to
thank N.Aizawa for his hospitality while visiting Osaka Prefecture University, where a great deal of
this work was undertaken. 
The authors thank the referees for valuable comments and suggestions.

%
%
%


\begin{thebibliography}{99}
%
%
\bibitem{Ree} R. Ree, ``Generalized Lie elements," \textit{Canad. J. Math.} \textbf{12}, 493 (1960).
\bibitem{rw1} V. Rittenberg and D. Wyler, ``Generalized Superalgebras," \textit{Nucl. Phys.} {\bf B 139}, 189 (1978) .
\bibitem{rw2} V. Rittenberg and D. Wyler, ``Sequences of $Z_2\otimes Z_2$ graded Lie algebras and superalgebras," \textit{J. Math. Phys.} {\bf 19}, 2193 (1978).
\bibitem{sch} M. Scheunert, ``Generalized Lie algebras," \textit{J. Math. Phys.} {\bf 20}, 712 (1979).
%
%
\bibitem{lr} J. Lukierski and V. Rittenberg, ``Color-De Sitter and Color-Conformal Superalgebras," \textit{Phys. Rev.} {\bf D 18},  385 (1978).
\bibitem{vas} M. A. Vasiliev, ``de Sitter supergravity with positive cosmological constant and generalized Lie superalgebras," \textit{Class. Quantum Grav.} {\bf 2},  645 (1985).
\bibitem{jyw} P. D. Jarvis, M. Yang and B. G. Wybourne, ``Generalized quasispin for supergroups",  \textit{J. Math. Phys.} {\bf 28},  1192 (1987).
\bibitem{zhe} A. A. Zheltukhin, ``Para-Grassmann extension of the Neveu-Schwartz-Ramond algebra", \textit{Theor. Math. Phys.} {\bf 71},  491  (1987) (\textit{Teor. Mat. Fiz.} {\bf 71},  218 (1987)).
\bibitem{Toro1} L. A. Wills-Toro, 
``$(I,q)$-graded Lie algebraic extensions of the Poincar\'e algebra, constraints on $I$ and $q$", \textit{J. Math. Phys.} \textbf{36}, 2085 (1995).
\bibitem{Toro2} L. A. Wills-Toro, 
``Trefoil symmetries I. Clover extensions beyond Coleman-Mandula theorem", \textit{J. Math. Phys.} \textbf{42}, 3915  (2001). 
\bibitem{tol2} V. N. Tolstoy,  ``Super-de Sitter and Alternative Super-Poincar\'e Symmetries," In: Dobrev V. (eds) \textit{Lie Theory and Its Applications in Physics. Springer Proceedings in Mathematics \& Statistics}, vol. 111, Springer, Tokyo, 2014.
\bibitem{tol} V. N. Tolstoy, ``Once more on parastatistics", \textit{Phys. Part. Nucl. Lett.} {\bf{11}},  933 (2014).
\bibitem{Bru} A. J. Bruce, ``On a $\mathbb{Z}_2^n$-graded version of supersymmetry," \textit{Symmetry}  \textbf{11},  116 (2019).
\bibitem{BruDup} A.J. Bruce and S. Duplij, ``Double-graded supersymmetric
	quantum mechanics,'' arXiv:1904.06975 [math-ph].	
\bibitem{NaAmaDoi} N. Aizawa, K. Amakawa and S. Doi, ``{\cal N}-Extension of
	double-graded supersymmetric and superconformal quantum mechanics,''
		\textit{J. Phys. A:Math. Theor.} \textbf{53}, 065205 (2020). 
		
\bibitem{Ueba} I. Ueba, ``Extended supersymmetry with central charges in Dirac
	action with curved extra dimensions,'' \textit{Phys. Rev. D} \textbf{100}, 105001 (2019). 
	
\bibitem{BruIbar} 
A. J. Bruce, E. Ibarguengoytia,
``The graded differential geometry of mixed symmetry tensors," \textit{Arch. Math.} (Brno) \textbf{55}, 123 (2019).	
	
\bibitem{aktt1} N. Aizawa, Z. Kuznetsova, H. Tanaka and F. Toppan, ``\Z2-graded Lie symmetries of the L\'evy-Leblond equations," \textit{Prog. Theor. Exp. Phys.} \textbf{2016}, 123A01 (2016).
\bibitem{aktt2} N. Aizawa, Z. Kuznetsova, H. Tanaka and F. Toppan, ``Generalized supersymmetry and L\'evy-Leblond equation", 
in S. Duarte \textit{et al} (eds), \textit{Physical and Mathematical Aspects of Symmetries.} Springer, Cham, 2017.
%
%
\bibitem{NAJS} N. Aizawa and J. Segar, ``\Z2 generalizations of ${\cal N} = 2$ super Schr\"odinger
algebras and their representations," \textit{J. Math. Phys.} \textbf{58}, 113501 (2017).
\bibitem{NaPsiJs} N. Aizawa, P. S. Isaac and J. Segar, ``$\mathbb{Z}_2 \times \mathbb{Z}_2$
generalizations of $ {\cal N}=1 $ superconformal Galilei algebras and their representations," \textit{J. Math. Phys.} \textbf{60}, 023507 (2019).
\bibitem{NA2} N. Aizawa, ``Generalization of superalgebras to color superalgebras and their
representations," \textit{Adv. Appl. Clifford Algebras} \textbf{28}, 28 (2018).
\bibitem{Barg} V. Bargmann, ``On unitary ray representations of continuous groups'', \textit{Ann. Math.} \textbf{54}, 1 (1954).
%
%
\bibitem{GFcocy}
I. M. Gelfand and D. B. Fuchs,  
``Cohomology of the Lie algebra of vector fields on the circle," 
\textit{Func. Anal. Appl.} \textbf{2}, 342 (1968).

\bibitem{Schtt}
M. Schottenloher, 
\textit{A Mathematical Introduction to Conformal Field Theroy},  
Springer, Berlin Heidelberg 2008.

\bibitem{ChaPre}
V. Chari and A. Pressley, 
``Unitary representations of the Virasoro algebra and a conjecture of Kac," 
\textit{Compos. Math.} \textbf{67}, 315 (1988).

\bibitem{GKO}
P. Goddard, A. Kent, D. Olive, 
``Unitary representations of the Virasoro and super-Virasoro algebras," 
\textit{Comm. Math. Phys.} \textbf{103}, 105 (1986).

\bibitem{FNZ}
D. B. Fairlie, J. Nuyts, C. K. Zachos, 
``A presentation for the Virasoro and super-Virasoro algebras," 
\textit{Comm. Math. Phys.} \textbf{117}, 595 (1988).
%
%
%
\bibitem{OvRo1}
V. Yu. Ovsienko and C. Roger, 
``Extensions of the Virasoro group and the Virasoro algebra by modules of tensor densities on $S^1$," 
\textit{Func. Anal. Appl.} \textbf{30}, 290 (1996).
\bibitem{OvRo2} 
V. Yu. Ovsienko and C. Roger, 
``Generalization of Virasoro group and Virasoro algebra through extensions by modules of tensor-densities on $S^1$," \textit{Indag. Mathem.} \textbf{9}, 277 (1998).
\bibitem{Mar}
P. Marcel, ``Extensions of the Neveu-Schwarz Lie superalgebra," 
\textit{Comm. Math. Phys.} \textbf{207}, 306 (1991).
\bibitem{MaOvRo}
P. Marcel, V Ovsienko, C. Roger, 
``Extension of the Virasoro and Neveu-Schwarz algebras and generalized Sturm-Liouville operators," 
\textit{Lett. Math. Phys.} \textbf{40}, 31 (1997).
%
%
\bibitem{BMS4} H. Bondi, M. G. J. van der Burg, A. W. K. Metzner, ``Gravitational Waves in General Relativity. VII. Waves from Axi-Symmetric Isolated Systems," \textit{Proc. Roy. Soc. Lond.}  \textbf{A 269}, 21 (1962). 
\bibitem{Sachs} R. K. Sachs, ``Gravitational Waves in General Relativity. VIII. Waves in Asymptotically Flat Space-Time," \textit{Proc. Roy. Soc. Lond.} \textbf{A 270}, 103 (1962).
%
%
\bibitem{GrJa} H. S. Green and P. D. Jarvis, ``Casimir invariants, characteristic identities, and Young diagrams for color algebras and superalgebras," \textit{J. Math. Phys.} \textbf{24}, 1681 (1983).
\bibitem{sch2} M. Scheunert, ``Graded tensor calculus", \textit{J. Math. Phys.} \textbf{24}, 2658 (1983).
\bibitem{sch3} M. Scheunert, ``Casimir elements of $\epsilon$-Lie algebras", \textit{J. Math. Phys.} \textbf{24}, 2671 (1983).
\bibitem{SchZha} M. Scheunert and R. B. Zhang, ``Cohomology of Lie superalgebras and their generalizations", \textit{J. Math. Phys.} \textbf{39}, 5024 (1998). 
\bibitem{Sil} S. D. Silvestrov, ``On the classification of 3-dimensional coloured Lie algebras," 
\textit{Banach Center Publ.} \textbf{40}, 159 (1997).
\bibitem{ChSiVO} X.-W. Chen, S. D. Silvestrov and F. Van Oystaeyen, ``Representations and cocycle twists of color Lie algebras", \textit{Algebr. Represent. Theor.} \textbf{9}, 633 (2006).

\bibitem{SigSil} G. Sigurdsson and S. D. Silvestrov, ``Bosonic realizations of the colour Heisenberg Lie algebras," . \textit{Nonlinear Math. Phys.} \textbf{13}, supplement  110 (2006).
\bibitem{PionSil} D. Piontkovski and S. D. Silvestrov, ``Cohomology of 3-dimensional color Lie algebras," \textit{J. Alg.} \textbf{316}, 499 (2007).
\bibitem{CART} R. Campoamor-Stursberg and M. Rausch de Traubenberg, ``Color Lie algebras and Lie algebras of order $F,$" 
\textit{J. Generalized Lie Theory Appl.} \textbf{3}, 113 (2009). 
\bibitem{NA1} N. Aizawa, ``Verma Modules over a $\mathbb{Z}_2 \otimes \mathbb{Z}_2$ graded superalgebra and invariant differential equations," \textit{Scientiae Mathematicae Japonicae}  \textbf{31}, 2018-4 (2018).
\bibitem{StoVDJ} N.I. Stoilova, J. Van der Jeugt, ``The \Z2-graded Lie superalgebra $ \mathfrak{pso}(2m+1|2n) $ and new parastatistics representations," \textit{J. Phys. \textbf{A}:Math. Theor.} \textbf{51}, 135201 (2018).
\bibitem{CGP1} T. Covolo, J. Grabowski and N. Poncin, ``$ \mathbb{Z}_2^n$-Supergeometry I: Manifolds and Morphisms," arXiv:1408.2755 [math.DG].
\bibitem{CGP2} T. Covolo, J. Grabowski and N. Poncin, ``$ \mathbb{Z}_2^n$-Supergeometry II: Batchelor-Gawedzki Theorem," arXiv:1408.2939 [math.DG].
\bibitem{CGP3} T. Covolo, J. Grabowski and N. Poncin, ``Splitting theorem for $ \mathbb{Z}_2^n$-supermanifolds," \textit{J. Geom. Phys.} \textbf{110}, 393 (2016).
\bibitem{CGP4} T. Covolo, J. Grabowski and N. Poncin, ``The category of $ \mathbb{Z}_2^n$-supermanifolds," \textit{J. Math. Phys.} \textbf{57}, 073503 (2016).
\bibitem{Pon} N. Poncin, ``Towards integration on colored supermanifolds," 
\textit{Banach Center Publ.} \textbf{110}, 201 (2016).
\bibitem{CoKPo} T. Covolo, S. Kwok and N. Poncin, ``Differential calculus on  $\mathbb{Z}_2^n$-supermanifolds," arXiv:1608.00949 [math.DG].
\bibitem{BruPon} A. J. Bruce and N. Poncin, ``Functional analytic issues in $ \mathbb{Z}_2^n$-geometry," arXiv:1807.11739 [math-ph].

\bibitem{BruPon2}
A. J. Bruce and N. Poncin,  ``Products in the category of-manifolds," 
\textit{J. Nonlinear Math. Phys.} \textbf{26}, 420 (2019).

\bibitem{MohSal} M. Mohammadi and H.  Salmasian, ``The Gelfand-Naimark-Segal construction for unitary representations of $\mathbb Z_2^n$-graded Lie supergroups," arXiv:1709.06546 [math-ph].
%
%
%
\bibitem{NAJS2} N. Aizawa and J. Segar, ``Aspects of infinite dimensional $\ell$-super Galilean conformal algebra," \textit{J. Math. Phys.} \textbf{57}, 123502 (2016).
%
%
\bibitem{FrScSo} L. Frappat, A. Sciarrino and P. Sorba, \textit{Dictionary on Lie Algebras and Superalgebras}, Academic Press, London 2000. 
%
%
%
%
%
%
%


\end{thebibliography}
\end{document}